\documentclass[twocolumn,secnumarabic,amssymb, superscriptaddress, aps, prb]{revtex4-2}
\usepackage{graphicx}
\usepackage[T1]{fontenc}
\usepackage{lmodern}
\usepackage{color}
\usepackage{natbib}
\usepackage{amsmath}
\usepackage{latexsym}
\usepackage{amsfonts}
\usepackage{subfigure}
\usepackage{picins}

\begin{document}

\title{Electronic subbands in the \emph{a}-LaAlO$_3$/KTaO$_3$ interface revealed by quantum oscillations in high magnetic fields}

\author{Km Rubi}
\email{rubi.km@ru.nl}
\affiliation{High Field Magnet Laboratory (HFML-EMFL) and Institute for Molecules and Materials, Radboud University, 6525 ED Nijmegen, The Netherlands}

\author{Shengwei Zeng}
\affiliation{Department of Physics, National University of Singapore, 117551 Singapore}

\author{Femke Bangma}
\affiliation{High Field Magnet Laboratory (HFML-EMFL) and Institute for Molecules and Materials, Radboud University, 6525 ED Nijmegen, The Netherlands}

\author{Michel Goiran}
\affiliation{Laboratoire National des Champs Magnétiques Intenses (LNCMI-EMFL), Université de Toulouse, CNRS, INSA, UPS, 143 Avenue de Rangueil, 31400 Toulouse, France}

\author{A. Ariando}
\email{ariando@nus.edu.sg}
\affiliation{Department of Physics, National University of Singapore, 117551 Singapore}

\author{Walter Escoffier}
\affiliation{Laboratoire National des Champs Magnétiques Intenses (LNCMI-EMFL), Université de Toulouse, CNRS, INSA, UPS, 143 Avenue de Rangueil, 31400 Toulouse, France}

\author{Uli Zeitler}
\affiliation{High Field Magnet Laboratory (HFML-EMFL) and Institute for Molecules and Materials, Radboud University, 6525 ED Nijmegen, The Netherlands}

\date{\today}

\begin{abstract}
Investigating Shubnikov-de Haas (SdH) oscillations in high magnetic fields, we experimentally infer the electronic band structure of the quasi-two-dimensional electron gas (2DEG) at the ionic-liquid gated amorphous (\emph{a})-LaAlO$_3$/KTaO$_3$ interface.
The angular dependence of SdH oscillations indicates a 2D confinement of a majority of electrons at the interface. 
However, additional SdH oscillations with an angle-independent frequency observed at high tilt angles indicate the coexistence of 3D charge carriers extending deep into the KTaO$_3$. 
The SdH oscillations measured in magnetic fields perpendicular to the interface show four frequencies corresponding to four 2D subbands with different effective masses (0.20 $m_e$ - 0.55 $m_e$). The single-frequency oscillations originating from 3D electrons yields a larger effective mass of $\sim$ 0.70 $m_e$.
Overall, the inferred subbands are in good agreement with the theoretical-calculations and angle-resolved photoemission spectroscopy studies on 2DEG at KTaO$_3$ surface.  
\end{abstract}

\maketitle

\section{Introduction}

Due to its fundamental complexity and potential applications in new-generation semiconductor and spintronic devices, the two-dimensional electron gas (2DEG) formed at the interface between two insulating transition metal oxides has triggered significant attention in the past decade \cite{hwang2012emergent, annurev-matsc, huang2018interface, noel2020non}. The 2DEG at perovskite oxide interfaces with strong spin-orbit coupling was first discovered for LaAlO$_3$/SrTiO$_3$ (LAO/STO) heterointerface \cite{ohtomo2004high, PhysRevLett.104.126803, PhysRevB.101.245114}. This interface exhibits several exceptional phenomena, for example, the coexistence of ferromagnetism and superconductivity \cite {li2011coexistence, bert2011direct} and sets up a benchmark for other perovskite oxide based interfaces. In recent years, the 2DEG of  even higher mobility and stronger spin-orbit coupling with electrons originating from the Ta:5d orbitals has been revealed at interfaces based on KTaO$_3$ (KTO), a band insulator ($E_g \sim$ 3.6 eV) and polar oxide \cite{AppPhysLett.10.1063, zhang2017highly, zhang2018high, wadehra2020planar}. Additionally, the 2D superconductivity with higher critical temperature has been reported at the KTO surface \cite{ueno2011discovery}  and KTO based interfaces \cite{liu2021two, chen2021two} of high carrier densities ($n \gtrsim 5 \times 10^{13}$ cm$^{-2}$). \\ 

From the perspective of electronic band structure, the KTO based 2DEG is  similar to the STO-2DEG where electrons  primarily occupy $t_{2g}$ orbitals of transition metal ions. More specifically, density functional theory (DFT) calculations for the LAO/KTO interface reveal a partial occupancy of  d$_{xy}$, d$_{xz}$, and d$_{yz}$ subbands arising from the t$_{2g}$ orbitals of Ta:5d band, and a dominant electronic contribution of the low-lying d$_{xy}$ subband  \cite {cooper2012enhanced, wang2016creating}. Tight-binding calculations for the 2DEG at the KTO surface predict a strong mixing of d$_{xy}$, d$_{xz}$, and d$_{yz}$ orbitals driven by spin-orbit coupling and a reconstruction of orbital symmetries of subbands \cite{PhysRevB.86.121107}. These findings are experimentally supported by angle-resolved photoemission spectroscopy (ARPES) measurements on oxygen-deficient KTO surface which mapped out parabolic and isotropic light ($m^* \sim 0.3 m_e$) and heavy ($m^* \sim 0.7 m_e$) subbands \cite{PhysRevLett.108.117602, PhysRevB.86.121107}. In addition, quantum transport experiments performed at $T=2$~K and up to $B=14$~T on metallic KTO surface, where oxygen vacancies were induced through Ar$^+$ irradiation, have revealed a single frequency Shubnikov-de Haas (SdH) oscillations assigned to heavy electrons subband ($m^* = 0.8 m_e$) \cite{harashima2013coexistence}.

In order to fully resolve the SdH oscillations corresponding to the closely spaced subbands of KTO-based 2DEGs, we have performed magneto-transport measurements on the amorphous ({\em{a}})-LAO/KTO interface in high magnetic fields (30 T continuous and 55 T pulsed). Since the large lattice constant of KTO restricts the epitaxial growth of a heterointerface, we create a conducting interface by growing an amorphous {\em{a}}-LAO thin film on KTO.  
High-mobility carriers are induced at the interface using ionic-liquid (IL) gating at room temperature as reported before for {\em{a}}-LAO/STO  \cite{zeng2016liquid, PhysRevLett.121.146802}. 
We measured two devices of different carrier densities, device-1 ($ n$ = 1.7 $\times 10^{13}$ cm$^{-2}$) and device-2 ($ n$ = 1.1 $\times 10^{13}$ cm$^{-2}$), in high magnetic fields. Large-amplitude SdH oscillations of several frequencies are observed for both devices. For an in-depth characterization of the SdH oscillations, we study their tilt-angle and temperature dependencies for device-1. The oscillations measured for different tilt-angles reveal the coexistence of 2D and 3D conduction channels. The analysis of temperature dependent SdH oscillations yields the effective masses ranging from $m^* \approx$ 0.2 $m_e$ to 0.55 $m_e$ and, therefore, confirms the occupancy of light and heavy subbands. Moreover, a back-gate dependent study on device-2 provides additional smaller amplitude and higher frequency oscillations, barely visible in the raw data, in the low-density regime.

\section{Experimental details}

Our \emph{a}-LAO/KTO devices consist of 4 nm LAO thin film deposited on the patterned Hall bar on the surface of a 500 $\mu$m thick KTO(001) substrate using pulsed laser deposition. During deposition, the substrate temperature was kept at 25$^o$ C and the oxygen partial pressure was  $p_{O_2} \sim 10^{-4}$ mbar. Six-terminal Hall bars of width 50 $\mu$m and length 160 $\mu$m between the longitudinal voltage leads were fabricated by conventional photolithography technique using AlN films as a hard mask, see Fig.~\ref{F1}(a) and (c).  The Hall bars were bonded with Al wires using ultrasonic wire bonder.
A silver-paint electrode, (G) in Fig.~\ref{F1}(a), not connected to the Hall-bar was used as a counter-electrode for ionic-liquid gating with N, N-diethyl-N-methyl-N-(2-methoxyethyl)ammonium bis (trifluoromethyl sulphonyl)imide (DEME-TFSI). 

Simultaneous measurements of longitudinal and Hall resistance were carried out in high dc magnetic field ($B_{max}$ = 30 T) at HFML-Nijmegen and high pulsed magnetic field ($B_{max}$ = 55 T with duration of 300 ms) at LNCMI-Toulouse. The dc-field measurements were performed in a $^3$He system  and a dilution refrigerator, and the pulsed field measurements in a $^3$He system. For pulsed field measurements, we apply a dc current excitation of 5 $\mu$A whereas we use a lower ac current of 100 nA for measurements in continuous fields. An in-situ rotation mechanism was used to change the angle between magnetic field orientation and the normal direction of the interface.

\section{Results and discussion}

\begin{figure}[!htp]
\includegraphics[width=8.9cm]{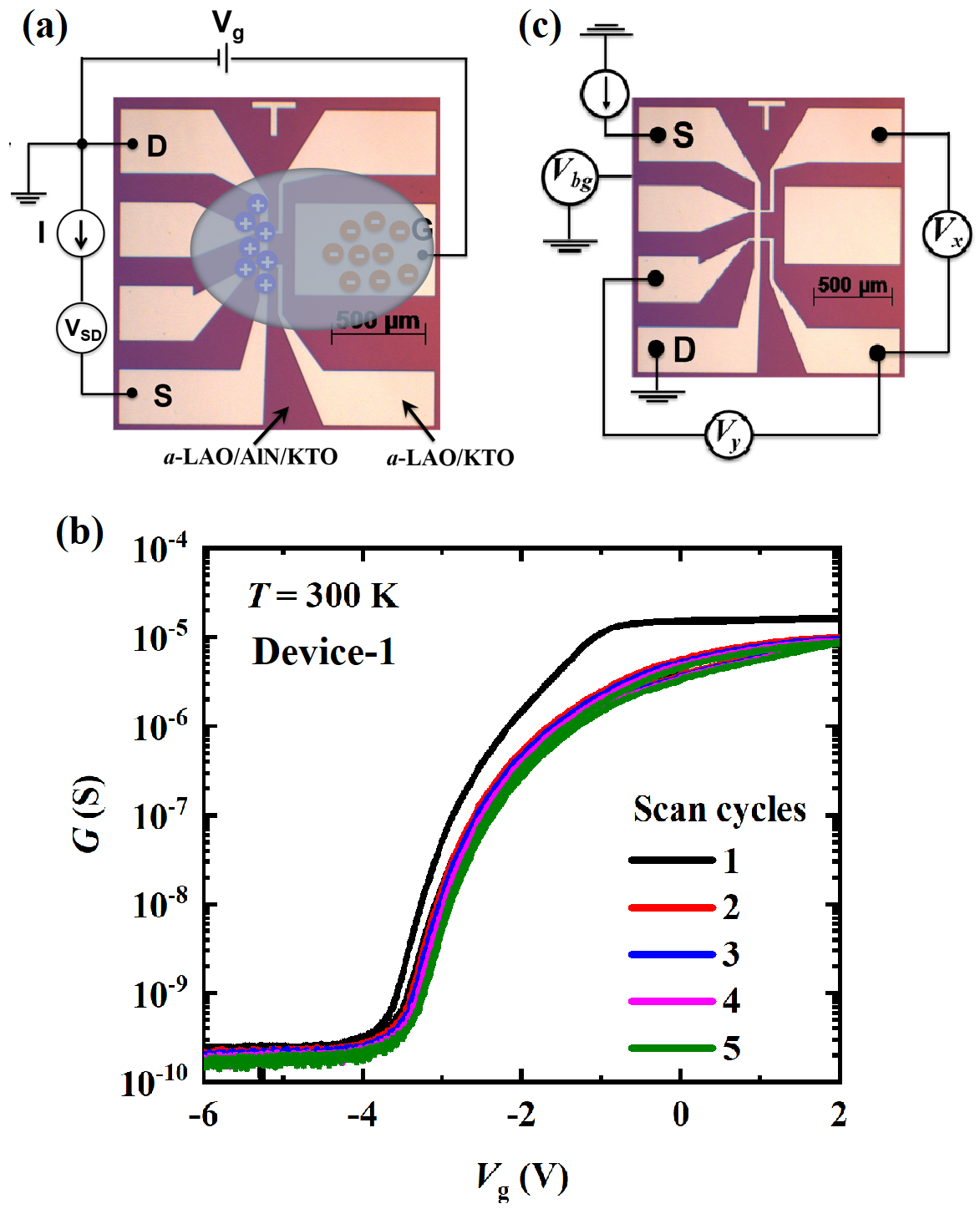} 
\caption{\label{F1} (a) Optical microscopy image of the Hall bar patterned \emph{a}-LAO/KTO device, including the gate (G), drain (D), and source (S) electrodes, the ionic liquid (IL) drop, and the electrical circuit used to bias and measure the device. (b) Two-probe conductance ($G = I_{SD}/V_{SD}$) of the device-1 as a function of the IL gate voltage ($V_g$) calculated from the measured current while applying  a source-drain voltage ($V_{SD}$). The device switches from a high-conductance to a low-conductance state when varying $V_g$ from positive to negative values. (c)  Transport measurement scheme after the IL gating process. The back-gate voltage $V_{bg}$ was applied across the KTO substrate.}
\end{figure}

Fig.~\ref{F1}(a) shows a top-view of the \emph{a}-LAO/KTO device including a measurement  scheme for the IL gating. A tiny drop of the IL covers the gate electrode and the conducting KTO channel. The source-drain current, $I_{SD}$ is measured with varying  IL gate voltage, $V_g$ at a constant source-drain voltage, $V_{SD}$ of 3 V, and we show the conductance ($G = I_{SD}/V_{SD}$) as a function of $V_g$ in Fig.~\ref{F1}(b). The device switches from a high-conductance ($ \sim 10^{-5}$ S) to a low-conductance ($ \sim 10^{-10}$ S) state when $V_g$ sweeps from +2 V to $-$6 V. A large hysteresis in $G(V_g)$ in the first scan cycle demonstrates an irreversible doping response, whereas the doping becomes almost reversible with a small hysteresis from the second cycle onwards. The onset of conduction and the hysteretic behavior can be explained by electromigration of oxygen ions from the \emph{a}-LAO  towards interface thereby irreversibly filling oxygen vacancies \cite{Jeong1402, li2013suppression, PhysRevLett.121.146802, leighton2019electrolyte}. After scanning for a few cycles, the gate sweep was stopped at $V_g$ = 0 V where the device remains stable even after the IL was cleaned off. The device was then cooled down to 2 K leaving a conducting channel at the \emph{a}-LAO/KTO interface with carrier density $n_H \sim 10^{13}$ cm$^{-2}$ and mobility, $\mu_H \sim 10^4$ cm$^2$V$^{-1}$s$^{-1}$, comparable to the values observed in IL-gated crystalline LAO/STO devices \cite{gallagher2015high, zeng2016liquid}. The transport measurments in high magnetic fields were carried out after a few months of device fabrication and IL gating treatment. The schematic of simultaneous measurements of longitudinal and Hall voltage is displayed in Fig. 1(c). By appying  back-gate voltage ($V_{bg}$) across the KTO substrate, we were able to adjust the electron concentration at the interface.

\begin{figure*}[!htp]
\includegraphics[width=6.8in]{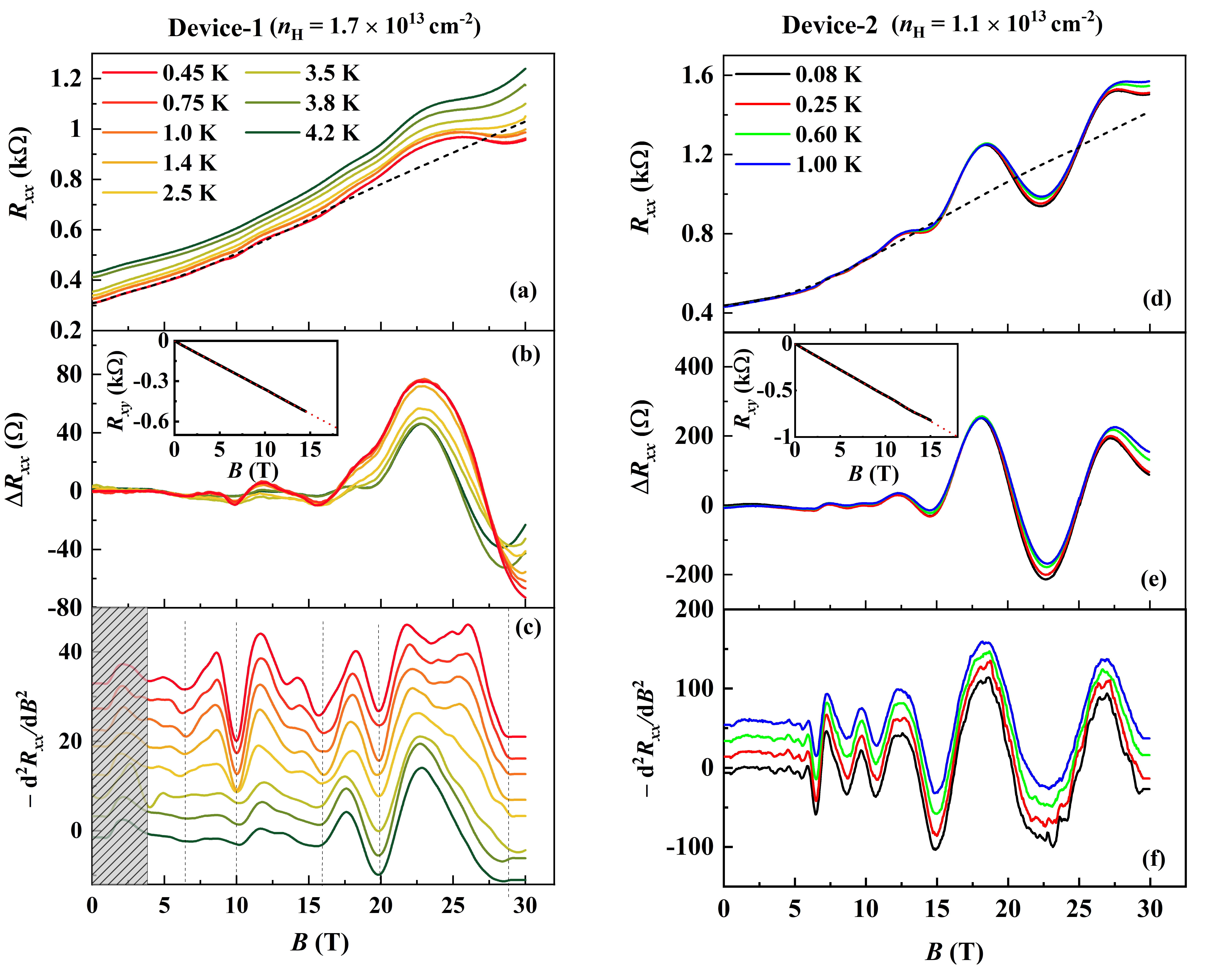}
\caption{\label{F2} {\bf Observation of quantum oscillations in magnetoresistance of \emph{a}-LAO/KTO devices.}
Left panels: Magnetic field dependence of transport data at different temperatures for device-1 (a) longitudinal resistance ($R_{xx}$), (b) oscillatory resistance ($\Delta R_{xx}$) after subtracting a polynomial background, and (c) the second-order derivative of $R_{xx}$ with respect to $B$ ($-$d$^{2}R_{xx}$/d$B^2$).
Right panels: Transport data at different temperatures for device-2 (d) $R_{xx}$, (e) $\Delta R_{xx}$, and (f) $-$d$^{2}R_{xx}$/d$B^2$. The dashed lines in (a) and (d) are the non-oscillating background at the lowest temperatures. The $-$d$^{2}R_{xx}$/d$B^2$ vs $B$ curves are shifted along the y-axis for clarity. 
Hall resistance data $R_{xy}$ ($B$)  are shown with linear fit (dashed line) in the insets of (b) device-1 and (e) device-2. The shaded region in Fig. \ref{F2}(c) covers a temperature-independent rapid increase in $R_{xx}(B)$ in low-fields which is not used in the analysis of SdH oscillations. }
\end{figure*}

\begin{figure*}[!htp]
\includegraphics[width=7.0 in]{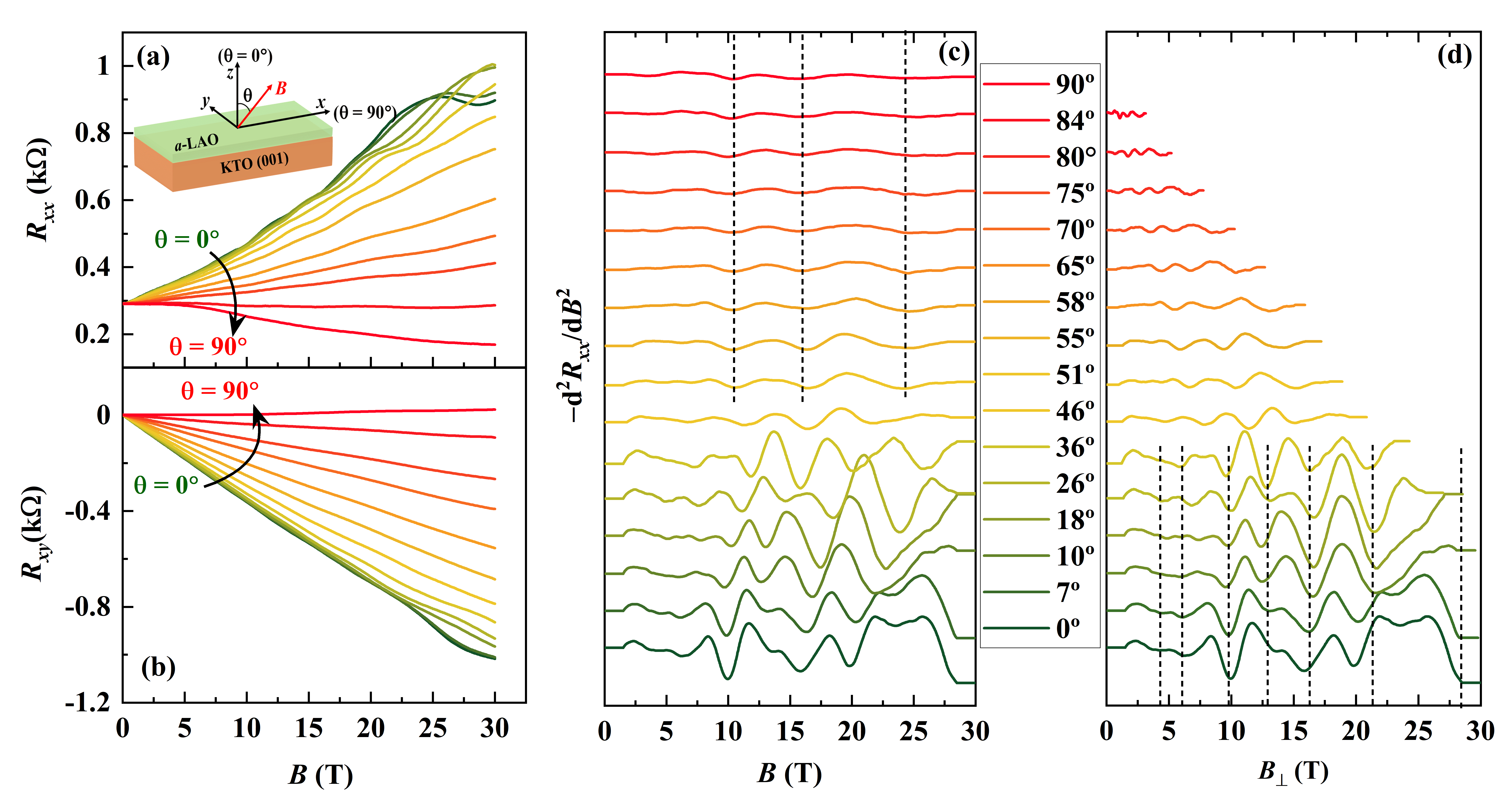} 
\caption{\label{F3} {\bf Angular dependence of the quantum oscillations in device-1.}
 (a) $R_{xx}$  and (b) $R_{xy}$  as a function of magnetic field for different angles $\theta$ = 0$^o\rightarrow 90^o$ at a  temperature $T$ = 400 mK. $\theta$ is the angle between the normal to KTO (001) plane and the magnetic field direction as depicted in inset (a).
Quantum oscillations in $R_{xx}(B)$ visualized in $-$d$^{2}R_{xx}$/d$B^2$ as a function of (c) the total magnetic field $B$ and (d) the perpendicular field component  $B_{\perp}$ = $B$cos$\theta$, respectively.  The curves are shifted along the y-axis for clarity. The vertical dashed lines mark oscillation minima as a guide to the eyes. The evolution of oscillations with tilt angles indicates the coexistence of 2D and 3D electronic states at the ionic liquid gated \emph{a}-LAO/KTO interface.}
\end{figure*}

 Fig.~\ref{F2}(a) shows the longitudinal resistance, $R_{xx}$ for device-1 measured at different temperatures in $^3$He system while sweeping magnetic field ($B$) oriented perpendicular to the interface. Additionally, the Hall resistance data, $R_{xy} (B)$ is displayed in the inset of Fig.~\ref{F2}(b). The device-1 exhibits Hall carrier density $n_H$ = $1.7 \times 10^{13}$ cm$^{-2}$ estimated from a linear fit of $R_{xy} (B)$ and mobility $\mu_H$ = 5310 cm$^2$V$^{-1}$s$^{-1}$. While the carrier density in the IL-gated \emph{a}-LAO/KTO devices remains almost same over a long period, we noticed a significant drop in the mobility after a few months of IL gating treatment. 
From Fig.~\ref{F2}(a), a positive magnetoresistance (MR) with clear quantum oscillations superimposed on it is observed. After subtracting a non-oscillating background resistance (dashed line in Fig.~\ref{F2}(a) for $T$ = 0.45 K) from the $R_{xx} (B)$ data, $\Delta R_{xx} (B)$ is shown in Fig.~\ref{F2}(b). The oscillations start developing at $\sim$ 4 T and, as expected, their amplitude increases with decreasing temperature and increasing magnetic field. For a clearer picture of these oscillations, we also show the second-order derivative of $R_{xx}$ ($-$d$^{2}R_{xx}$/d$B^2$) as a function of $B$ in Fig. 2(c). 
The $-$d$^{2}R_{xx}$/d$B^2(B)$ curve at 4.2 K shows few oscillations, whereas more oscillations pop-up when lowering the temperature. This effect originates from the quantization of density of states of a spin-split band or multiple subbands.

In order to further resolve these oscillations, we have measured device-2 with $n_H$ = $1.1 \times 10^{13}$ cm$^{-2}$ and $\mu_H$ = 5800 cm$^2$V$^{-1}$s$^{-1}$ in a dilution refrigerator at temperatures down to 80 mK. $R_{xx}(B)$ and $\Delta R_{xx}(B)$ at different temperatures are shown in Fig. 2(d) and 2(e), respectively. Due to the higher mobility and lower temperatures, more resolved and larger amplitude oscillations are observed for device-2, which start developing from $B$ $\sim$ 3 T as revealed by the $-$d$^{2}R_{xx}$/d$B^2$ data shown in Fig.~\ref{F2}(f). Additionally, high-frequency (HF) and small-amplitude oscillations are superimposed on usual low-frequency and large-amplitude oscillations at $T$ = 80 mK and 300 mK, which will be discussed in detail later. 
 
In addition to the data shown in the left panels of Fig. 2, device-1 was measured at different tilt angles ranging from 0$^{\circ}$ to $90^{\circ}$  at a base temperature of 400 mK in the $^3$He system. The tilt angle $\theta$, as illustrated in the schematic in Fig. 3(a), is measured between the magnetic field $B$ and the normal to the interface.  
For the in-plane field orientation ($\theta$ = 90$^{\circ}$), $B$ is parallel to the current. Fig.~\ref{F3}(a) and (b) show $R_{xx}(B)$ and $R_{xy}(B)$, respectively, measured while varying the field in positive direction. To confirm the data-symmetry, the measurements for $\theta$ = 0$^{\circ}$ and 90$^{\circ}$ were also carried out in negative field directions. As shown in the supplementary information S1, $R_{xy}(B)$ for $\theta$ = 0$^{\circ}$ is nearly symmetric with respect to zero field. However,  $R_{xx}(B)$ exhibits a large asymmetry which indicates an inhomogeneous distribution of electrons. 
We observed a large anisotropic behavior of $R_{xx}(B)$ with varying $\theta$. The positive MR decays with increasing $\theta$ and switches to negative MR at high tilt angles. Interestingly, a large and non-saturating negative MR ($\sim - 40 \%$ at $B$ = 30 T) is perceived for $\theta$ = 90$^{\circ}$. Such a negative in-plane MR strongly dependent on temperature and carrier density has also been reported for the 2DEG at the LAO/STO interface \cite{PhysRevLett.115.016803, MingYang2016, PhysRevB.80.140403, PhysRevB.84.075312} and has been atrributed to spin-dependent scattering.
The strong spin-orbit interaction ($\sim$ 0.47 eV) \cite{zhang2017highly, wadehra2020planar} accompanied with spin-disorder scattering then accounts for a large negative MR in the 2DEG at the $a$-LAO/KTO interface.

\begin{figure*}[!htp]
\includegraphics[width=7.2in]{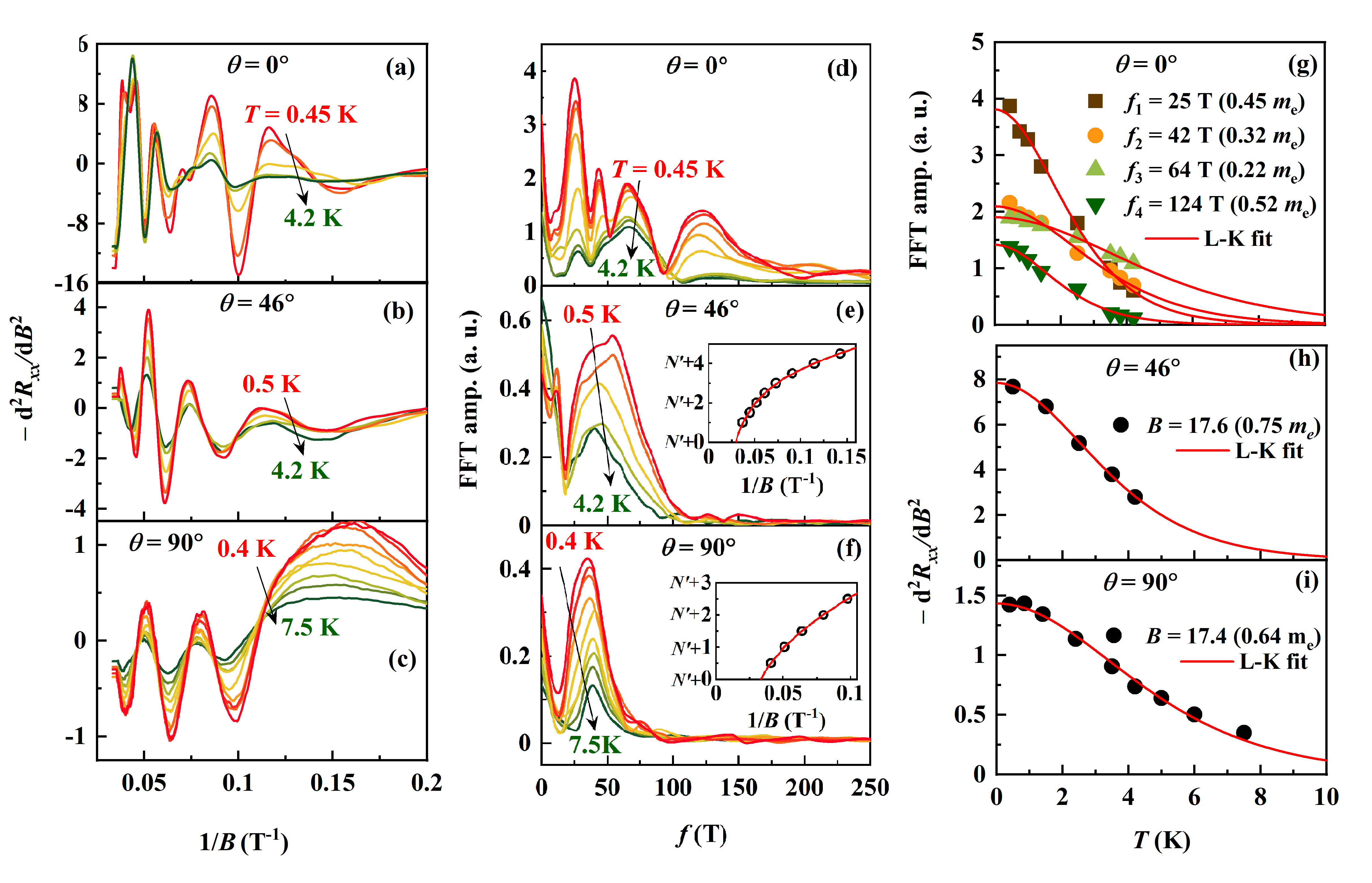} 
\caption{\label{F4}  {\bf Oscillations' frequency and cyclotron mass analysis for device-1.} Left column: inverse field dependence of the second-order derivative of $R_{xx}$ ($-$d$^{2}R_{xx}$/d$B^2$) at various temperatures and in different device's orientations (a) $\theta$ = 0$^{\circ}$, (b) $\theta$ = 46$^{\circ}$, and (c) $\theta$ = 90$^{\circ}$. The broad oscillation in the low-field regime for $\theta$ = 90$^{\circ}$ should not be counted as a SdH oscillation since it develops from an abrupt change in the negative MR around $\sim$ 6 T. Middle column: fast Fourier transform (FFT) analysis of $-$d$^{2}R_{xx}$/d$B^2$ versus $1/B$ data shown in the left column. FFT amplitude as a function of  frequency for (d) $\theta$ = 0$^{\circ}$, (e) $\theta$ = 46$^{\circ}$, and (f) $\theta$ = 90$^{\circ}$. Insets in (e) and (f) show the Landau plots (Landau level index ($N$) versus $1/B$) where maxima and minima of oscillations are assigned to the integer and half-integer indices, respectively, and $N'$ is an integer offset. Right column: (g) FFT amplitude as a function of temperature fitted with L-K equation for $\theta$ = 0$^{\circ}$. The peak-to-valley difference for a given oscillations' period versus temperature, $T$ plots fitted with L-K equation for (h) $\theta$ = 46$^{\circ}$, and (i) $\theta$ = 90$^{\circ}$. Symbols are the experimental data and solid-lines are the L-K fit. To avoid a crowded and unclear picture, we show oscillations only for few selected temperatures in (a); however, the FFT analysis for all temperatures is displayed in (d).}
\end{figure*}

To evaluate the dimensionality of the electronic states at the interface, we have examined the angular dependence of the SdH oscillations. Fig. 3 (c) shows the oscillations for different tilt angles displayed in the second-order derivative of $R_{xx}(B)$. The oscillations show a complex angle dependent evolution which can be attributed to the presence of several electronic sub-bands \cite {RN51}. For low tilt angles,  $\theta$ =0$^{\circ}$ ...  36$^{\circ}$, most of the oscillation minima progressively shift to higher magnetic fields. More specifically, their position depends on the perpendicular field component $B_{\perp}$ (= $B$cos$\theta$) only, see Fig.~\ref{F3}(d), which indicates the 2D confinement of conduction electrons at the interface. A slight mismatch in the cos$\theta$ scaling and the complex evolution of the oscillations' amplitude with $\theta$ (appearance and disappearance of some oscillations) most-likely results from the anticrossing of Landau levels from neighboring 2D sub-bands. In contrast, as depicted by the dashed lines in Fig.~\ref{F3}(c), the small-amplitude oscillations which are still visible at higher angles do not depend on the tilt angle, i.e.~they scale with the {\sl total} magnetic field. This observation suggests the existence of an additional 3D conduction channel parallel to the 2D channels at the interface. 

In order to characterize the nature of the observed sub-bands in more detail, the SdH oscillations for device-1 at different temperatures are analyzed for three specific angles $\theta$ = 0$^{\circ}$, 46$^{\circ}$, and 90$^{\circ}$ in Fig.~\ref{F4}. In the left panels, Fig.~\ref{F4}(a)-(c),  we show the SdH oscillations as a function of the inverse magnetic field  for the three different tilt angles. The corresponding fast Fourier transforms (FFTs) are then displayed in the middle panels, Fig.~\ref{F4}(d)-(f). We can observe four main frequencies -- $f_1$ = 25 T, $f_2$ = 42 T, $f_3$ = 64 T, and $f_4$ = 124 T -- in the FFT spectra at $\theta$ = 0$^{\circ}$ corresponding to four electronic subbands. The total 2D carrier density can be estimated from these frequencies  as $ n_{SdH}^{2D} = 2\frac{e}{h}\sum_{i} f_i $, where the factor 2 allows for the spin degeneracy. 
Using the above frequency values, we find $n_{SdH}^{2D} \approx 1.2\times 10^{13}$ cm$^{-2}$ slightly lower than the $n_{H}$ (1.7 $\times 10^{13}$ cm$^{-2}$) for this device. 

In comparison to $\theta$ = 0$^{\circ}$, fewer oscillations are observed for $\theta$ = 46$^{\circ}$ and 90$^{\circ}$ in the full field range, see Figs.~\ref{F4}(b) and (c), and the FFT spectrum is dominated by a single broad peak at $f \approx 38$~T. The single broad peak in FFT is manifested from non $1/B$-periodic oscillations which frequency monotonically increases with magnetic field. The Landau plots (Landau level index versus $1/B$) displayed in the inset of Figs. 4(e) and (f) specify $1/B$-aperiodic oscillations, similar to that observed in the 2DEG in LAO/STO \cite{MingYang2016, rubi2020aperiodic} and $\delta$-doped STO \cite{PhysRevB.82.081103}. As mentioned above, these oscillations are originated from a single 3D band in which the chemical potential or charge carrier density varies as the magnetic field increases \cite{rubi2020aperiodic}. 
Assuming a spherical Fermi surface, the average 3D carrier density can then be estimated from the SdH frequency $f$ as:
\newline 
$ n_{SdH}^{3D} = \frac{8}{3\sqrt{\pi}}\left(\frac{ef}{h}\right)^{3/2} = 1.3  \times 10^{18}~{\rm cm}^{-3}$

We extract the effective thickness of 3D conduction channel, $t_{3D}$ by comparing the $n_{SdH}^{3D}$ and 3D Hall carrier density, $n_{H}^{3D}$. As the $n_{SdH}^{2D}$ is a fraction of total carrier density ($n_H$ = $1.7 \times 10^{13}$ cm$^{-2}$), the remaining carriers ($n_{H}$ - $n_{SdH}^{2D}$ = $5 \times 10^{12}$ cm$^{-2}$) would be participating in the 3D oscillations. We calculate $t_{3D} \approx$ 40 nm as $n_{H}^{3D} = \frac{n_{H} - n_{SdH}^{2D}}{t_{3D}} = n_{SdH}^{3D}$. \\

Having established that a majority of electrons are confined at the interface, next we focus on cyclotron mass, $m_c$ analysis. To estimate the $m_c$ values, a fit of the temperature-dependent oscillations' amplitude to the standard Lifshitz-Kosevich (L-K) formula is employed \cite{shoenberg2009magnetic}.  
However, for multiple subbands systems in which oscillations of various frequencies are observed, the $m_c$ evaluated from the FFT amplitude is more reliable. Fig. 4(g) shows temperature dependence of FFT amplitude fitted with L-K formula 
\begin{equation}
A(T) = A_0 \frac {2{\pi}^2k_Bm_cT/\hbar eB_{eff}}{Sinh(2{\pi}^2k_Bm_cT/\hbar eB_{eff})} 
\end{equation}

where $\frac{1}{B_{eff}}= \frac{\frac{1}{B_{min}}+\frac{1}{B_{max}}}{2}$. 

The best fitting yields the $m_c$ values, 0.45 $m_e$ (25 T), 0.32 $m_e$ (42 T), 0.22 $m_e$ (64 T) and 0.52 $m_e$ (124 T), lower than those for 2DEG at LAO/STO interface \cite{RN51}.  A large error in the $m_c$ could be expected since the relatively large field window (5-30 T) is used in FFT analysis. We estimate up to 12$\%$ error from the $\Delta R_{xx}$ data simulated in the same field range using comparable values of frequencies and $m_c$ (supplementary information S2). Because the oscillations for $\theta$ = 46$^{\circ}$ and 90$^{\circ}$ are originated from a single band, the $m_c$ is calculated using the peak-to-valley difference for a given oscillations period. The fitted $m_c$ values, 0.75 $\pm$ 0.05  $m_e$ for  $\theta$ = 46$^{\circ}$ and 0.64 $\pm$ 0.06  $m_e$ for 90$^{\circ}$, yield an average $m_c \approx 0.7$ $m_e$ for the electrons in the 3D band.

\begin{figure}[!htp]
\includegraphics[width=8 cm]{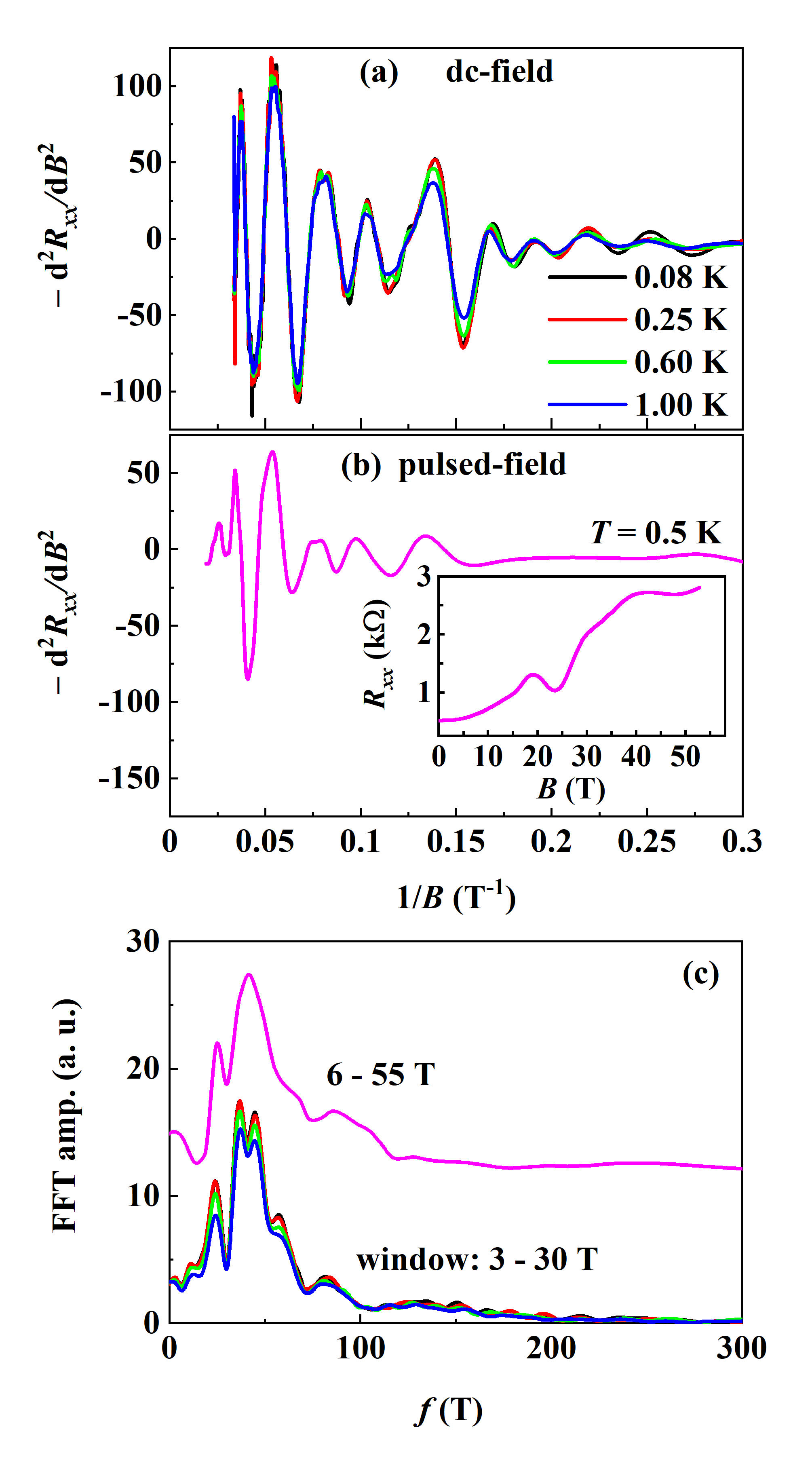} 
\caption{\label{F5} {\bf Analysis of SdH oscillations for device-2 measured in dc ($B$ = 0-30 T) and pulsed fields ($B$ = 0-55 T) applied perpendicular to the interface.} Main panels: Inverse field dependence of the second order derivative of the $R_{xx}$ ($-d^2R_{xx}/dB^2$) in (a) dc-field and (b) pulsed-field. Inset of (b) displays the $R_{xx}$ as a function of magnetic field. (c) A comparison of the FFT analysis in different windows of dc-field (3-30 T) and pulsed field (6-55T). The FFT plot for pulsed-field is shifted along y-axis for better comparison.}
\end{figure}

To add further validation on the value of the 2D subbands in {\em{a}}LAO/KTO interface, the quantum oscillations observed for device-2 in a perpendicular field orientation (right panel in Fig. 2) are analysed. Additionally, another experiment was carried out in pulsed magnetic field to extend the magnetic field range up to 55 T. The inset of Fig. 5(b) shows the $R_{xx}(B)$ data measured in the pulsed-field at $T$ = 500 mK. The inverse field dependence of the oscillations for dc and pulsed fields are shown in the main panel of Fig. 5(a) and (b), respectively. The results in the pulsed-field are similar to the dc-field data, although a higher current (leading to higher effective temperature) leads to a loss of resolution. Fig. 5(c) compares the FFT analysis of oscillations in dc and pulsed fields. Five main frequencies 23 T, 35 T, 43 T, 58 T and 82 T are observed from the FFT analysis of dc-field data. 
The small-amplitude 3D oscillations, which were not visible for device-1 when measured in the dc-field oriented perpendicular to the interface, are well resolved in the device-2 in the same field direction. 
Interestingly, $n_{SdH}^{2D}$ = $1.15 \times 10^{13}$ cm$^{-2}$ estimated from five frequencies of SdH oscillations is comparable to $n_{H}$, 1.1 $\times 10^{13}$ cm$^{-2}$. 

 \begin{figure}[!btp]
\includegraphics[width=8 cm]{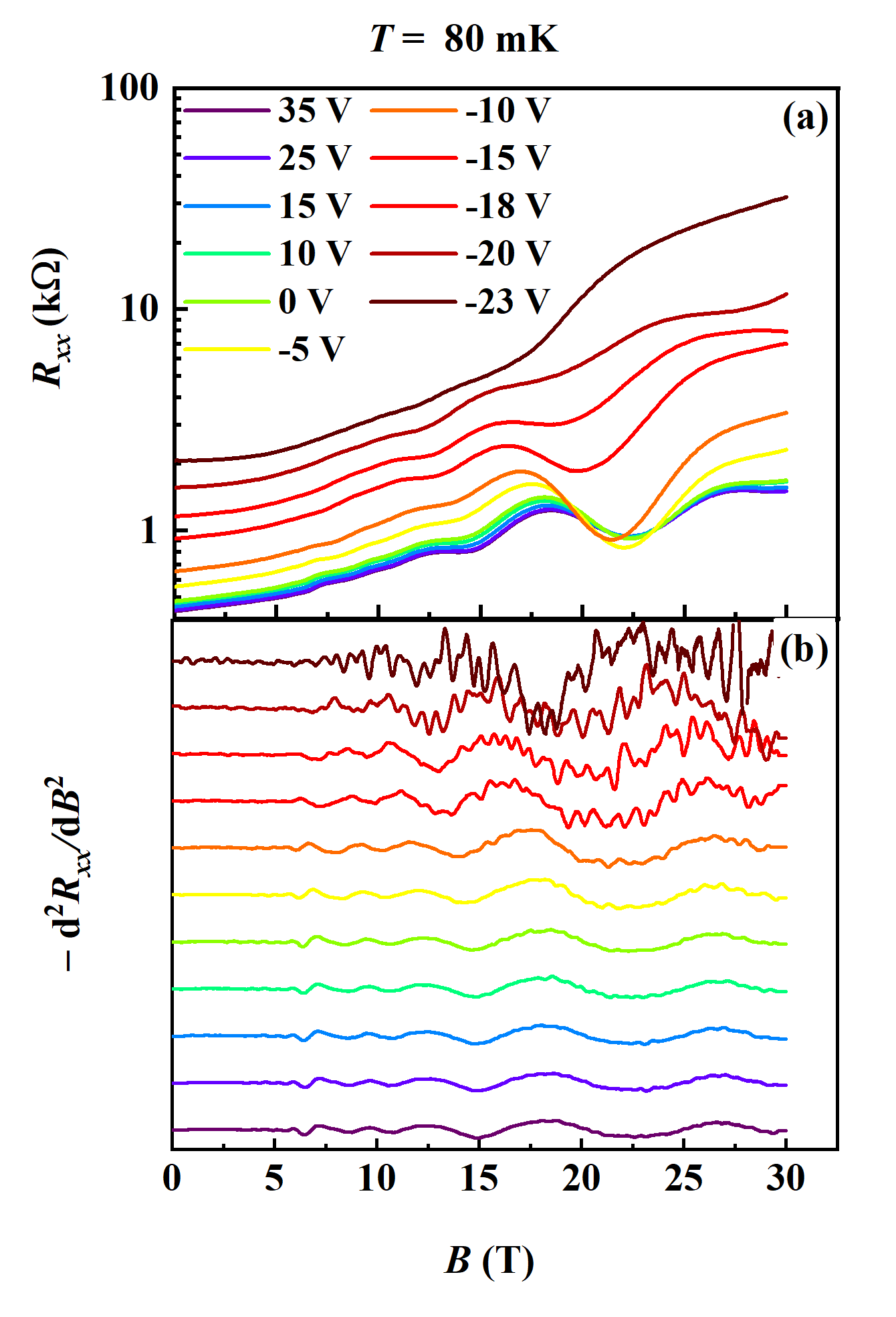} 
\caption{\label{F6}  {\bf Evolution of quantum oscillations with back gate voltage ($V_{bg}$) for device-2}. Magnetic field dependence of (a) $R_{xx}$ and (b) $-$d$^{2}R_{xx}$/d$B^2$  for different $V_{bg}$ and at $T$ = 80 mK. Oscillations spectra in (b) are vertically shifted for clarity. High-frequency oscillations superimposing the low-frequency SdH oscillations are clearly observed for negative values of $V_{bg}$.}
\end{figure}

Theoretically, the spin-orbit coupling in KTO leads to strong hybridization of $d_{xy}$, $d_{xz}$ and $d_{yz}$ orbitals and generates light and heavy mass subbands. In addition, the combination of spin-orbit coupling and 2D confinement in KTO based 2DEG leads to reconstruction of the orbital symmetries of 2D subbands and considerable renormalization of their masses \cite{PhysRevB.86.121107}. From tight-binding calculations, the effective masses of the light and heavy subbands along $\Gamma$-X axis are 0.25 $m_e$ and 0.5 $m_e$, respectively \cite{PhysRevB.86.121107}. The ARPES measurements on 2DEG at KTO surface confirms the light and heavy subbands of masses comparable with those predicted theoretically \cite{PhysRevLett.108.117602, PhysRevB.86.121107}. Our analysis on 2D oscillations concludes the occupancy of two light ($m^*$ = 0.22 - 0.32 $m_e$) and two heavy ($m^*$ = 0.45 - 0.52 $m_e$) subbands, which is in good agreements with ARPES and theoretical calculations. Moreover, the mass ($\approx 0.7 m_e$) estimated for 3D band is in very good agreement to the mass of the heavy subband of the bulk KTO \cite{PhysRevB.86.121107}.

Finally, we add a remark on another interesting observation mentioned earlier, the observation of HF oscillations in device-2 at very low temperatures (Fig. 2(f)). These oscillations are also investigated at different back-gate voltage, $V_{bg}$ applied through KTO substrate at $T$ = 80 mK. The variation in $R_{xx}(B)$ at different $V_{bg}$, +35 V $\rightarrow$ $-$23 V, is displayed in Fig. 6(a). The evolution of quantum oscillations with $V_{bg}$ is well-visible in $R_{xx}(B)$ curves and clearer in the oscillations spectra plotted as $-$d$^{2}R_{xx}$/d$B^2 (B)$ in Fig. 6(b). As expected, the SdH oscillations shift to lower field when $V_{bg}$ varies from +35~V to $-$23 V. Surprisingly, the HF oscillations' amplitude grows rapidly with increasing value of negative $V_{bg}$. Although the deep analysis of HF oscillations is beyond the scope of this paper, we examine the characteristic of these oscillations in brief. Interestingly, the HF oscillations are reproducible at different temperatures and field sweep rates (Supplementary information S3). We noticed the HF oscillations are not periodic in $1/B$ nor in $B$. The amplitude of oscillations, which is lower than the conductance quantum ($G_0 = 2e^2/h$ = 0.77 $\mu$S) at 80 mK and $V_g$  = $-$ 20 V, decays with increasing temperature (Fig. S3 in supplementary information). Most likely, these aperiodic and reproducible HF oscillations are resistance fluctuations originating from the quantum-interference of electron transport. Depopulating the electrons at the \emph{a}-LAO/KTO interface could lead to inhomogeneous 2DEG and form incoherent conductance channels. Such quantum interference phenomena have also been observed in the patterned 2DEG at the STO surface \cite{stanwyck2013universal} and the LAO/STO interface \cite{irvin2019strong}.

\section{Summary}

In summary, we have investigated transport properties of the ionic-liquid gated \emph{a}-LAO/KTO interface in high magnetic fields and at low temperatures. The SdH oscillations measured for low tilt angles reveal the two-dimensional nature of the electron gas at the interface. Additionally, the angle-independent oscillations observed in high tilt angles indicates the coexistence of 3D conduction channel of width $t_{3D}$ $\sim$ 40 nm. Observation of multiple frequencies of 2D oscillations confirms the occupancy of multiple subbands with light and heavy electrons, in good agreement with ARPES and theoretical calculations for KTO based 2DEG. The single-frequency 3D oscillations estimate the effective mass of $\sim 0.7 m_e$ corresponding to the heavy band. Furthermore, small-amplitude and high-frequency oscillations superimposing the low-frequency SdH oscillations emerge in the low carrier-density regime. Apart from fine observation electronic properties of the KTO-2DEG, this study demonstrates a possible route of studying quantum transport in 2DEG based on 5d-oxides for which creating an epitaxial heterointerface and thereby attaining the high-mobility electron gas is difficult. \\

\textbf{Acknowledgements}

We acknowledge the support of HFML-RU/FOM and LNCMI-CNRS, members of the European Magnetic Field Laboratory (EMFL). This study has been partially supported through the EUR grant NanoX n$^o$ANR-17-EURE-0009 in the framework of the “Programme des Investissements d’Avenir”. 
S.Z. and A.A. are supported by the Agency for Science, Technology, and Research (A*STAR) under its Advanced Manufacturing and Engineering (AME) Individual Research Grant (IRG) (A1983c0034), the National University of Singapore (NUS) Academic Research Fund (AcRF Tier 1 Grants No. R-144-000-391-114 and No. R-144-000-403-114), and the Singapore National Research Foundation (NRF) under the Competitive Research Programs (CRP Grant No. NRF-CRP15-2015-01).\\

%\large {\textbf{References}}

%\bibliographystyle{apsrev4-2}
%\bibliography{MS_LAO-KTO_PRR}

%

\newpage

\onecolumngrid

\section{Supplementary information}

\renewcommand{\section}{\textbf{S}}

\renewcommand\thefigure{S\arabic{figure}}

\setcounter{figure}{0}

\section{\textbf{1 Measurements in positive and negative field directions}}

\begin{figure*}[!htp]
\includegraphics[width=7 in]{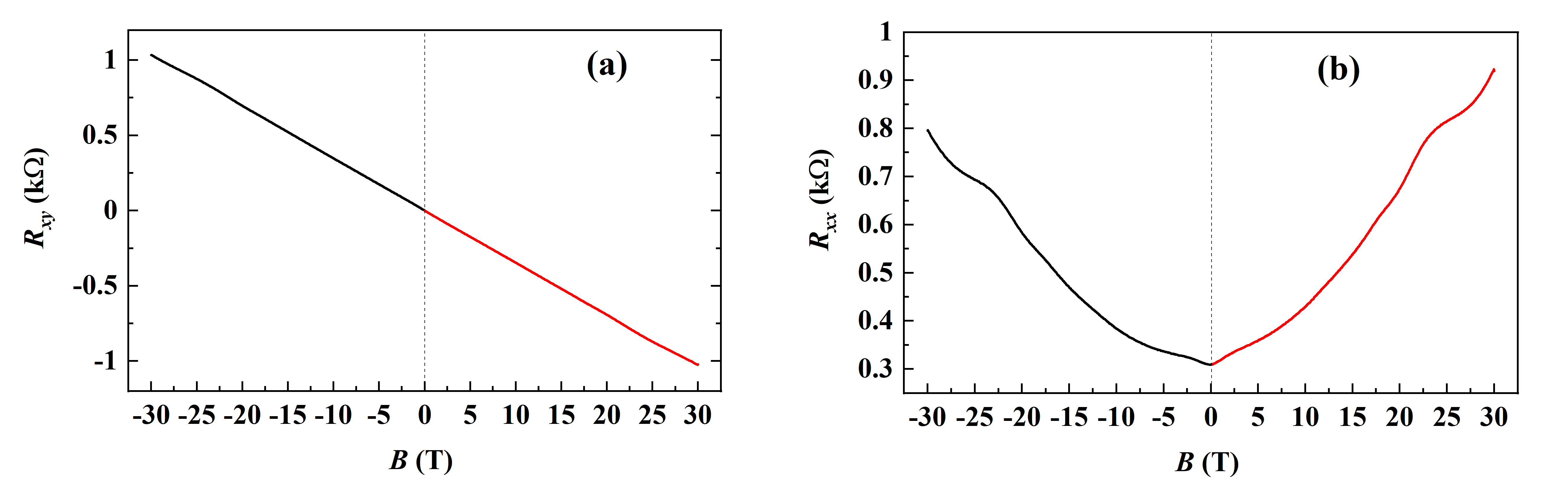}
\caption{Comparison of the transport data in both positive and negative field directions for $\theta$ = 0$^{\circ}$. Magnetic field dependence of (a) Hall resistance, $R_{xy}$ and (b) longitudinal resistance, $R_{xx}$. The $R_{xy}(B)$ is symmetric with respect to zero, however, a large asymmetry is observed for $R_{xx}(B)$. The periodicity of the oscillations is independent of the magnetic-field direction.}
\label{figS1} 
\end{figure*}

\section{\textbf{2 Cyclotron mass analysis from the FFT of simulated oscillations}}

\begin{figure*}[!htp]
\includegraphics[width=5.5 in]{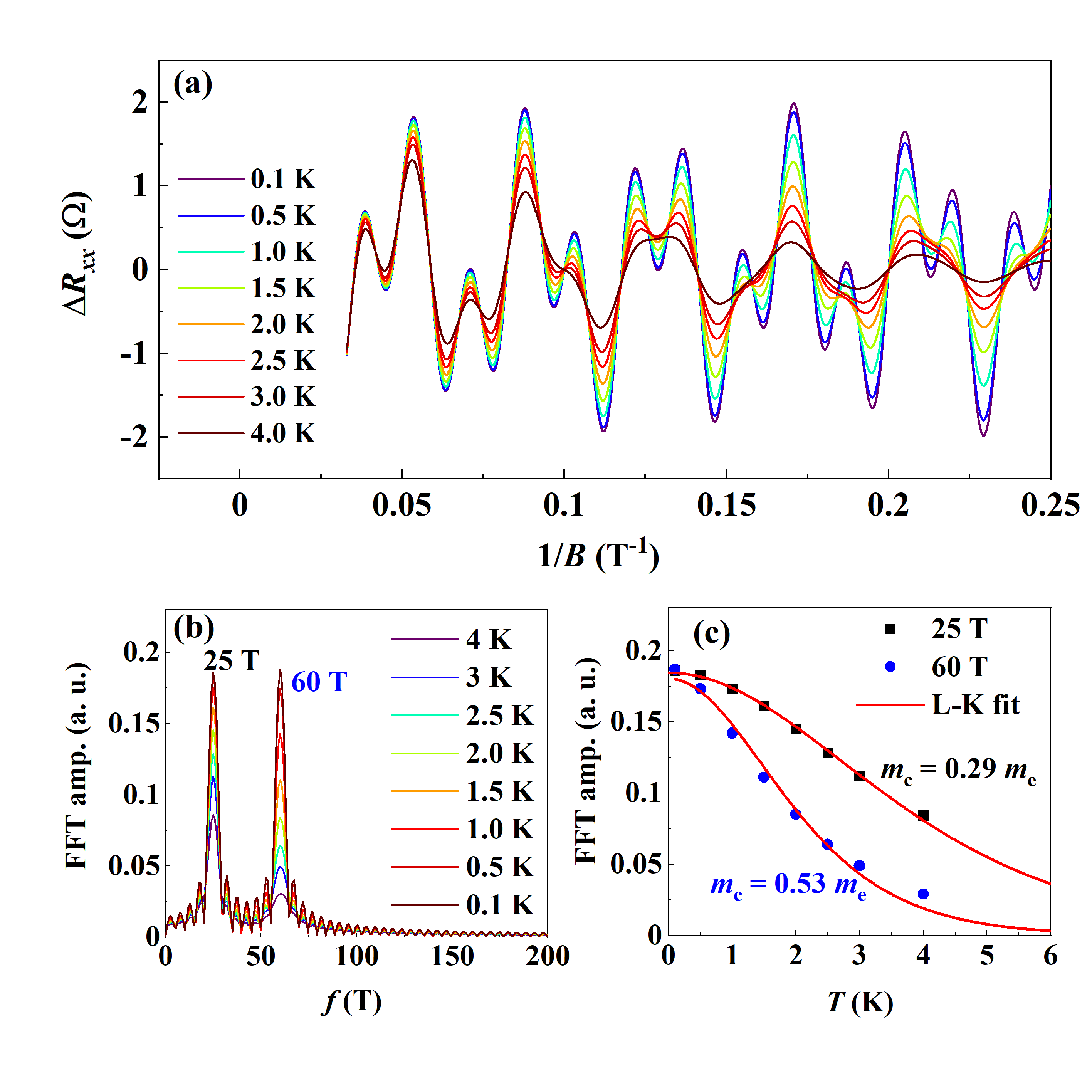}
\caption{\label{Fig1} (a) Inverse field dependence of the simulated SdH oscillations at different temperatures. (b) FFT analysis of the oscillations, and (c) FFT amplitude fit to the Eq. $ A(T) = A_0 \frac {2{\pi}^2k_Bm_cT/\hbar eB_{eff}}{Sinh(2{\pi}^2k_Bm_cT/\hbar eB_{eff})}$, where $\frac{1}{B_{eff}}= \frac{\frac{1}{B_{min}}+\frac{1}{B_{max}}}{2}$.}
\label{figS2} 
\end{figure*}

In order to quantify the error in the cyclotron masses estimated from the temperature dependent FFT amplitude, we simulate the oscillations in a field range of 4 - 30 T. The Lifshitz-Kosevich expression we used to simulate the SdH oscillations is 
\begin{equation} \label{eq:1}
\Delta R_{xx} = \sum_{i} exp(2\pi^2k_Bm_{ci} T_{Di}/\hbar eB)\frac {2{\pi}^2k_Bm_{ci}T/\hbar eB}{Sinh(2{\pi}^2k_Bm_{ci}T/\hbar eB)} sin\left(\frac{2\pi f_i}{B}\right)
\end{equation} 
where  $f_i$, $m_{ci}$ and $T_{Di}$ are the frequency of the oscillations, cyclotron mass and Dingle temperature, respectively, corresponding to the i$_{th}$ subband. 

For simplicity, we simulated the oscillations considering only two subband of different properties. The frequency and cyclotron mass values are deliberately chosen close to the values obtained experimentally for 2DEG at $\emph{a}$-LaAlO$_3$/KTaO$_3$. For $i$ = 1, $f$ = 25 T, $m_c$ = 0.3 $m_e$ and $T_D$ = 3.0 K; and for $i$ = 2, $f$ = 60 T, $m_c$ = 0.6 $m_e$ and $T_D$ = 3.5 K. \\
The inverse field dependence of the oscillations and corresponding FFT amplitude as a function of frequency are shown in Fig. S2(a) and S2(b), respectively. The FFT analysis produces main peaks in FFT amplitude exactly at the input frequencies 25 T and 60 T. The L-K fitting of the temperature dependent FFT amplitude (Fig. S2(c)) estimates the $m_c$ values, 0.29 $m_e$ and 0.53 $m_e$ corresponding to the 25 T and 60 T oscillations. The error in the   $m_c$ values is calculated as $\sim 4\%$ for light ($m_c$ = 0.3 $m_e$) and  $\sim 12\%$ for heavy ($m_c$ = 0.6 $m_e$) masses.
We noticed that the high frequency oscillations, for example $f_1$ = 1000 and $f_2$=1500 T, also gives the similar error even if FFT is performed in a smaller window.\\ 
%We beleive that the FFT analysis which is a mathematical expression  

\newpage

\section{\textbf{3 High-frequency (HF) oscillations}}

\begin{figure*}[!htp]
\includegraphics[width=9 cm]{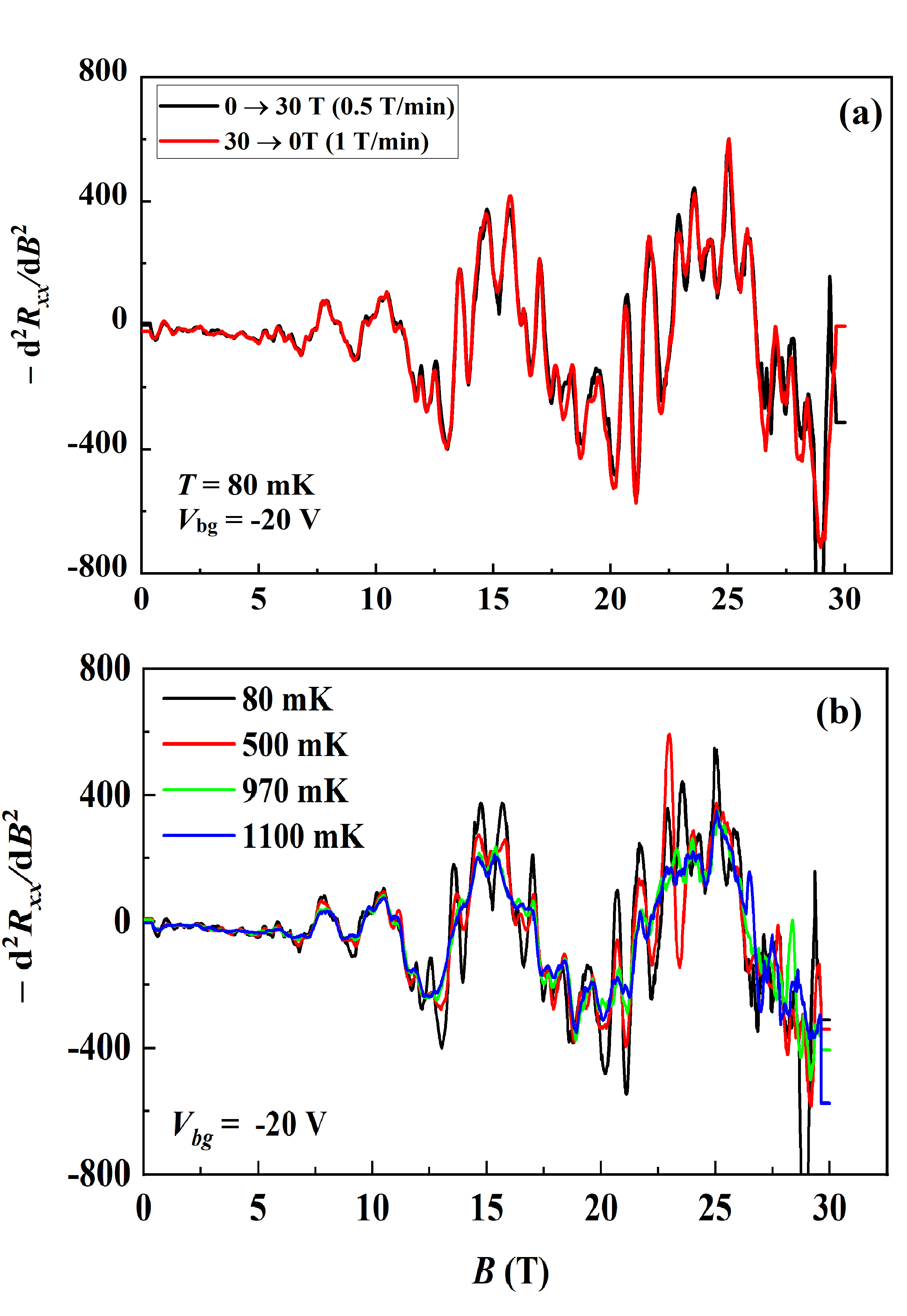}
\caption{\label{Fig1} Reproducibility of HF oscillations. Magnetic field dependence of the second order derivative of the longitudinal resistance $R_{xx}$ (a) for different field sweep rates 0.5 T/min and 1T/min at $T$ = 80 mK, and (b) at different temperatures for $V_g$ = $-20$ V. The high-frequency oscillations are robust and reproducible.}
\end{figure*}

To check the reproducibility of HF oscillations, we measured $R_{xx}$ at different field sweep rates (0.5 and 1 T/min) at the lowest possible temperature $T$ = 80 mK and at different temperatures with fixed sweep rate (1T/min). We show the oscillations spectra at different sweep rate and different temperatures for $V_{bg} = - 20$ V in Fig. S3(a) and (b), respectively. The HF oscillations are well-reproducible with varying the field sweep rate or temperature. As expected, increasing temperature suppresses the oscillations' amplitude. 

Coexistance of low and high frequency oscillations makes analysis very difficult. Because the HF oscillations' amplitude is much smaller in comparison to the LF oscillations, a second-order polynomial smoothing  of the raw data within a very small point-window erases all HF oscillations. After subtracting the smoothed data from the raw data, we plot oscillations in the form of oscillating conductance ($\Delta G$ = 1/$\Delta R$) as a function of $B$ in Fig. S4. 

\begin{figure}[h]
    
    {\begin{minipage}[c]{10cm} \includegraphics[width=9cm]{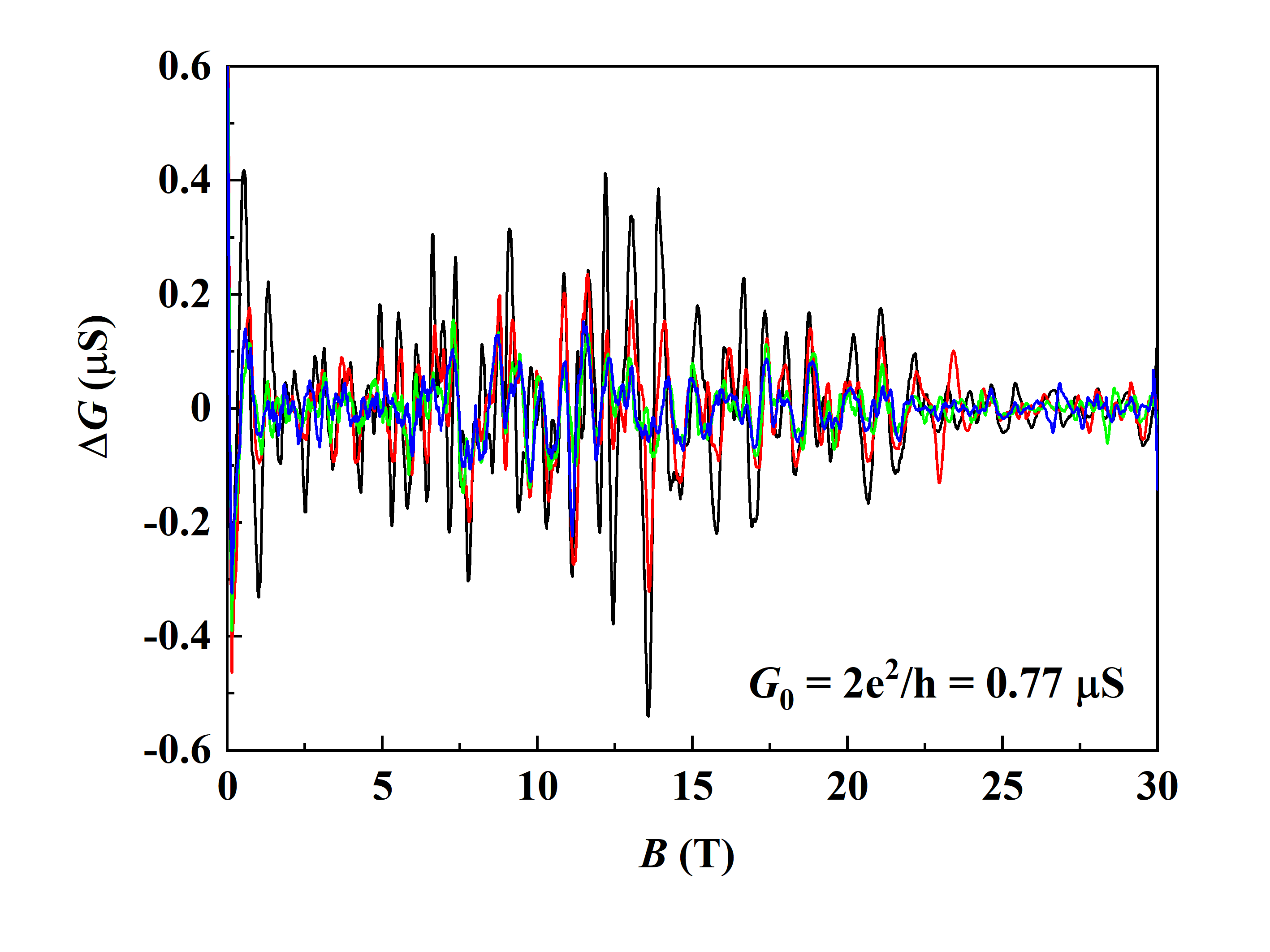} \end{minipage}
\begin{minipage}[c]{6cm} \caption{\label{Fig.S1_1} Oscillating conductance, $\Delta G$ calculated after subtracting the smoothed $R_{xx}(B)$ data from the raw data. The conductance remains lower than the value of quantum conductance ($2e^2/h = 0.77 \mu$S)} \end{minipage}
}
\end{figure}

\end{document}